\documentclass[12pt,preprint]{aastex}
\usepackage{graphicx}
\usepackage{amsmath}
\usepackage{longtable}
\usepackage{color}

\newcommand{\be}{\begin{equation}}
\newcommand{\ee}{\end{equation}}
\newcommand{\bea}{\begin{eqnarray}}
\newcommand{\eea}{\end{eqnarray}}

\newcommand{\etal}{et al.}

\newcommand{\td}{3^{\circ}}
\newcommand{\xo}{\overline{x}}

\newcommand{\pht}{\tilde{\phi}_1}
\newcommand{\bdelta}{\mbox{\boldmath$\delta$}}

\begin{document}


\bibliographystyle{apsrev}

\title{Embedded Lensing Time Delays, the Fermat Potential, and the Integrated Sachs-Wolfe Effect}

\author{Bin Chen\altaffilmark{1,2}, Ronald Kantowski\altaffilmark{1}, Xinyu Dai\altaffilmark{1}}

\altaffiltext{1}{Homer L. Dodge Department of Physics and Astronomy,
University of Oklahoma, Norman, OK 73019, USA}

\altaffiltext{2}{Research Computing Center, Department of Scientific Computing,
Florida State University, Tallahassee, FL 32306, USA, bchen3@fsu.edu}


\begin{abstract}
We derive the Fermat potential for a spherically symmetric lens embedded in an FLRW cosmology and use it to investigate the late-time integrated Sachs-Wolfe (ISW) effect, i.e., secondary temperature fluctuations in the cosmic microwave background (CMB) caused by individual large scale clusters and voids.
We present a simple analytical expression for the temperature fluctuation in the CMB across such a lens as a derivative of the lens' Fermat potential.
This formalism is applicable to both linear and nonlinear density evolution scenarios,  to arbitrarily large density contrasts, and to all open and closed background cosmologies.
It is much simpler to use and makes the same predictions as  conventional approaches.
In this approach the total temperature fluctuation can be split into a time-delay part and an evolutionary part.
Both parts must be included for cosmic structures that evolve and both can be equally important.
We present very simple ISW models for cosmic voids and galaxy clusters to illustrate the ease of use of our formalism.
We use the Fermat potentials of simple cosmic void models to compare predicted ISW effects with those recently extracted from WMAP and \emph{Planck} data by stacking large cosmic voids  using the aperture photometry method. 
If voids in the local universe with large density contrasts are no longer evolving we find that the time delay contribution alone predicts values consistent with the measurements.
However, we find that for voids still evolving linearly, the evolutionary contribution cancels a significant part of the time delay contribution and results in predicted signals that are much smaller than recently observed.
\end{abstract}

\keywords{cosmic background radiation---cosmology: theory---gravitation---gravitational lensing: strong---large-scale structure of universe}

\maketitle

\section{Introduction}

The cosmic microwave background (CMB) exhibits an intrinsic temperature anisotropy caused by the primordial density perturbations.
Determining the spectrum of this primary anisotropy is of fundamental importance to modern cosmology.
The accurate evaluation of this spectrum is complicated by the presence of {\it secondary} anisotropies caused by the thermo/kinetic Sunyaev-Zeldovich effect \citep{Zeldovich69,Sunyaev80,Birkinshaw99}, the integrated Sachs-Wolfe (ISW) effect \citep{Sachs67}, gravitational lensing \citep{Cole89,Seljak96a}, and reionization \citep{Vishniac87,Hu00,Miralda00}.
In this paper we are mainly interested in the ISW effect caused by inhomogeneities that exist along the line of sight from observation back to the CMB's decoupling at around redshift $z\sim1100$ (the last scattering surface).
ISW fluctuations caused by nonlinear growths of density perturbations  at low redshifts  were investigated in \cite{Rees68} using Swiss cheese models \citep{Einstein45, Schucking54} and their presence is referred to as the Rees-Sciama (RS) effect.
The importance of the ISW/RS effect to modern cosmology is multifold.
First, studying the effect probes the  evolution history of cosmic structures and the dynamics of dark energy.
For example, in a flat $\Lambda$CDM universe, a nonzero cosmological constant causes the expansion of the universe to accelerate, resulting in a decay of the gravitational potentials associated with cosmic voids, and thus leaving its evolutionary imprint on the observed CMB anisotropies.
Detecting this effect is considered a crucial test for the presence of dark energy, since it directly probes the negative pressure nature of dark energy and complements other methods which are geometrical in nature, e.g., using SNe\,Ia as standard candles \citep{Riess98,Perlmutter99} or using baryonic acoustic oscillations as standard rulers \citep{Eisenstein05}.
Second, the ISW/RS effect caused by local inhomogeneities, in particular, cosmic voids, has been used extensively to explain several large angle anomalies in CMB anisotropies observed by the Wilkinson Microwave Anisotropy Probe (WMAP; Bennett et al.\ 2003).
For example, the octopole planarity (a preferred axis along which the octopole power is suppressed), the alignment between the quadrupole and the octopole moments, the anomalously cold spot on the south hemisphere, and the temperature asymmetry in the large angle power between opposite hemispheres all have possible explanations via the ISW/RS effect \citep{Tegmark03,Velva04,Cooray05, Inoue06, Rudnick07, Szapudi14a, Szapudi14b,Finelli14, Nadathur14a}.
Accurately measuring the ISW/RS effect is also important for detecting possible non-Gaussian signatures in the primordial density spectrum.
To measure the primordial non-Gaussianity of the initial density spectrum, {\it secondary} non-Gaussianities caused by the ISW/RS effect, have to be systematically investigated and removed \citep{Kim13}.
For example, the ISW-lensing bispectrum has been recently detected by \emph{Planck}
\citep{Planck14a}.

Even though \cite{Rees68} pointed out that the time delay experienced by a CMB photon crossing a mass over-density can cause significant temperature fluctuations, the RS effect today usually refers only to those fluctuations caused by time-dependent potentials.
Detailed theoretical investigation of the ISW  effect continues to date \citep{Dyer76,Nottale84,Martinez90,Panek92,Seljak96b,Cooray02a, Cooray02b, Sakai08,Valkenburg09,Smith09}. 
Observational detection of the ISW effect  includes the two following approaches.
First by constructing the correlation function between the CMB temperature sky map and tracers of large scale structures in different wavelengths \citep{Boughn04,Ho08,Dupe11}, hot and cold spot directions are compared with galaxy over and under-densities.
This correlation has been tentatively found by the \cite{Planck14b}.
The second approach is based on aperture photometry, i.e.,  patches of CMB maps around known large cosmic voids and clusters are stacked and averaged in order to minimize contamination from {\it primordial} anisotropies and to detect  cold or hot spots due to the ISW effect.
Recently several groups claimed detection of the ISW effect using this second method \citep{Granett08a,Granett08b,Planck14b}, although doubts exist \citep{Hernandez10,Nadathur12,Ilic13} based on linear structure growth rates in a $\Lambda$CDM universe.
One major controversy surrounding the void-stacking method is the large value of the observed fluctuations compared to theoretical predictions and numerical simulations \citep{Maturi07,Cai10,Cai14, Nadathur12,Ilic13,Hernandez13, Flender13, Watson14,Hotchkiss15}.
Another surprising result of this method is the strange shape of the photometric profile, including the hot ring in the outer part of the profile \citep{Ilic13,Planck14b}.
This structure is also hard to explain within the framework of the linear ISW effect in a standard $\Lambda$CDM cosmology (see Fig.~9 of Planck Collaboration et al.\ 2014b).

Here we present a theory which  is easy to use and which may help to resolve these differences.
We investigate the effect of lensing by individual large scale structures on observed CMB temperature fluctuations using our recently developed embedded  lens theory \citep{Kantowski10,Kantowski12,Kantowski13,Kantowski14,Chen10,Chen11,Chen13}.
Our lens is the same as a general Swiss cheese condensation but so named to emphasize the consequences of embedding, i.e., the consequences of making the lens mass a contributor to the mean mass density of the universe.
An embedded lens is constructed by first removing a comoving sphere from a homogeneous Friedman-Lema\^itre-Robertson-Walker (FLRW) universe, producing a vacuole, or a Swiss cheese void,
then deforming that mass into an arbitrary spherical shape, and  placing it back into the void in such a way as to keep Einstein's equations satisfied
throughout \citep{Einstein45, Schucking54}.\footnote{The notion of a Swiss cheese void should not be confused with a cosmic void. Cosmic voids are regions under-dense in gravitating matter and are often identified by looking for regions under-dense in galaxies.}
Deformed mass densities can be over-dense in the central region and represent  galaxy clusters or under-dense in the central region and represent  cosmic voids.
Because of general relativity's (GR) boundary conditions the average density throughout the refilled Swiss cheese void is essentially the same as the background cosmology.
Consequently the dense central region of an embedded cluster lens model must be surrounded by an under-dense (void like) region (see Fig.~\ref{fig:cheese}).
For the same reason an embedded lens model for a cosmic void which is low in central mass density must be surrounded by an over-dense region (often a shell of mass).
By using the embedded lens model and satisfying Einstein's equations we are assured that gravitational approximation errors are not inherent at any order.
Until now most analytical expressions for temperature fluctuations in the CMB caused by crossing individual cosmic structures have been complicated and were best presented in graphical form. By keeping only the lowest order lensing terms and using the Fermat potential we are able to give a simple analytical expression for such fluctuations.

In Section \ref{sec:theory} we present the fundamental theoretical result of this paper, Eq.~(\ref{calT2}), a simple relation between the Fermat potential and the ISW effect.
In Section \ref{sec:examples} we apply this theory to two simple embedded lens models, one for linearly growing structures and one for co-expanding structures.
For simplicity we assume the clusters have stopped evolving relative to the background cosmology but for the voids we additionally compute an evolutionary correction by assuming the voids are still evolving linearly.
Conclusions are drawn in Section \ref{sec:conclusions}.
In Appendix~\ref{sec:appendix} we present a subtle derivation of Eq.~(\ref{calT2}).
In  Appendix~\ref{append:static} we check the validity  of Eq.~(\ref{calT2}) by comparing its prediction for the embedded point mass lens with the prediction made by the conventional approach.

\section{The Fermat Potential of Embedded Spherical Lenses and the ISW Effect}\label{sec:theory}

\begin{figure*}
\includegraphics[width=0.9\textwidth,height=0.18\textheight]{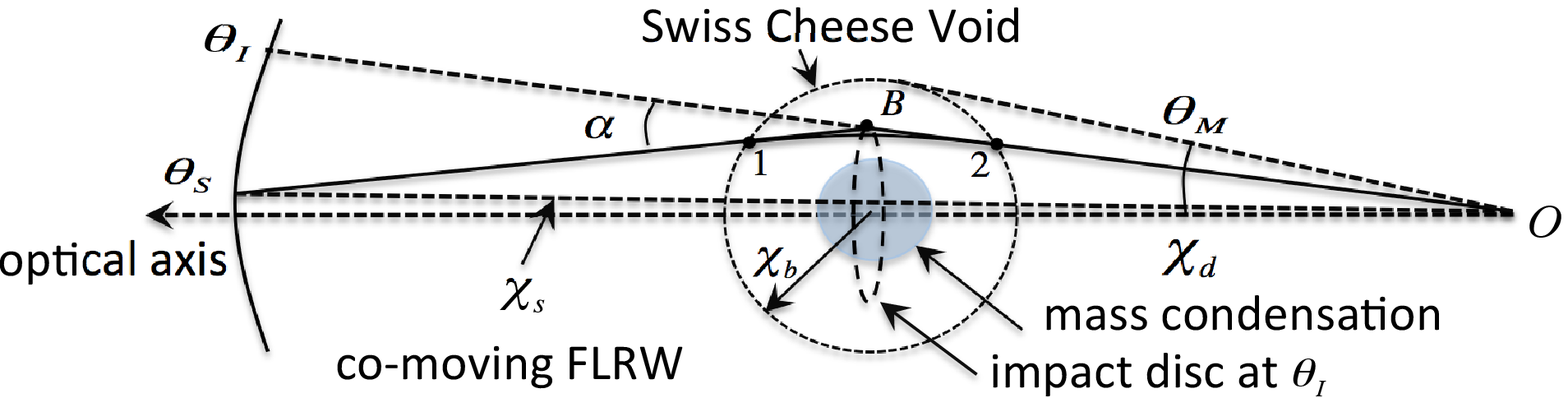}
\caption{ The comoving geometry of an embedded cluster lens.
	$\theta_S$ and $\theta_I$ are the source and image angles, respectively;
	$\chi_d$ and $\chi_s$ are respective comoving distances of the lens and the source.
	The (constant) angular size of the void is $\theta_M\equiv \chi_b/\chi_d$ where $\chi_b$ is the comoving radius of the void whose physical radius is $r_d=R(t_d)\chi_b$ at cosmic time $t_d$ defined by the deflector's redshift $1+z_d=R_0/R(t_d)$.
	The shadowed area represents the extended spherically symmetric mass condensation  of the cluster.
	The dashed circle shows the impact disc of angular radius $\theta_I$ used to compute the projected mass fraction $f$ used in Eqs.\,(\ref{spherical})-(\ref{fdelta}).
        The equivalent figure for a void has a mass condensation surrounding a low density central region and an outward deflection angle $\alpha$.
}
\label{fig:cheese}
\end{figure*}

The geometry of a CMB photon crossing an embedded lens of angular radius $\theta_M$ is sketched in Fig.~\ref{fig:cheese}.
We assume a standard FLRW background universe with radius $R(t)$ in which last scattering occurs well before the CMB photons encounter the cluster or void of interest.
In the homogeneous universe a CMB photon's frequency evolves with time according to $\nu_{\td}(t)= \nu_{\td}\,R(t_0)/R(t),$ reaching us at time $t_0$ with frequency $\nu_{\td}$.
If this CMB photon is gravitationally lensed by an inhomogeneity at redshift $z_d$ before reaching us, its frequency evolution is altered in its passage through the lens between points 1 and 2 (see Fig.~\ref{fig:cheese}).
It enters the inhomogeneity at cosmic time $t_1$ with CMB frequency  $\nu_{\td}(t_1)$ and exits at time $t_2$ but with a frequency $\nu_{\rm sc}(t_2)$, not equal to the CMB frequency $\nu_{\td}(t_2)$, and then evolves  again in the homogeneous FLRW background according to $\nu_{\rm sc}(t)= \nu_{\rm sc}\,R(t_0)/R(t)$, reaching us with frequency $\nu_{\rm sc}\ne \nu_{\td}$.
Because all frequencies are similarly affected the temperature of the deflected photons will differ (after exiting)  from those undeflected
by
\be
\frac{\Delta{\cal T}}{{\cal T}}=\frac{\nu_{\rm sc}-\nu_{\td}}{\nu_{\td}}.
\label{calT1}
\ee

It is well known in conventional gravitational lensing theory that the arrival time at the observer of a photon emitted by a background source will be delayed if it is lensed by a foreground mass condensation, and that this time delay can be written as the sum of a geometrical and a potential part, $T=T_g+T_p$ \citep{Cooke75}.
In conventional lensing theory time delay refers to the difference in arrival times of two separate lens images.
In this work the time delay $T$ refers to the delayed arrival time of a single lensed image compared to its would be arrival time in the absence of the lens.
The difference in the $T$'s for two separate images gives their conventional arrival time difference. In all cases the time delays are of order $r_s/c$ (the time it takes light to cross the Schwarzschild radius of the lens) and are small compared to the age of the universe.
Within the conventional lensing theory the total time delay $T$ we are introducing is proportional to the Fermat potential whose minimization, according to Fermat's least time principle  (i.e., $\delta T/\delta \theta_I =0$ where $\theta_I$ the image angle, see Fig.~\ref{fig:cheese}), gives the lens equation \citep{Schneider85,Blandford86}.
In \cite{Kantowski13} we rigorously derived the lens equation for the embedded point mass lens and generalized that equation, at first order, to any spherically symmetric embedded lens
\be\label{spherical}
\theta_S=\theta_I-\frac{\theta_E^2}{\theta_I}\left[f(\theta_I/\theta_M,z_d)-f_{\rm RW}(\theta_I/\theta_M)\right],
\ee
where $\theta_S$ and $\theta_I$ are the source and image angles, $\theta_E=\sqrt{2r_sD_{ds}/D_dD_s}$ is the standard Einstein ring angle ($D_d,$ $D_s$ and $D_{ds}$ are respectively the angular diameter distances of the lens, the source, and the source with respect to the lens),
\be
f_{\rm RW}(x)\equiv 1-\left[\sqrt{1-x^2}\right]^3,\hskip .5in 0\le x\le 1,
\label{fRW}
\ee
is the projected mass fraction of the homogeneous sphere removed to form the Swiss cheese void ($x\equiv \theta_I/\theta_M$ is the normalized image angle),  and $f(x,z_d)\equiv M(x\,\theta_M,z_d)/M(\theta_M)$ is the fraction of the lens' mass projected within the impact disk defined by $\theta_I$ at the lens redshift $z_d$ (schematically shown in Fig.\,\ref{fig:cheese}).
For an inhomogeneity evolving at a rate different from the background FRLW universe, $f(x,z_d)$ depends on the cosmic time or equivalently the lens redshift $z_d.$
We have rigorously derived the time-delay function, i.e., the Fermat potential for the embedded point mass lens (see Eq.\,(7) in Kantowski et al.\ 2013).
We now generalize that first order result and give the  Fermat potential  for any spherically symmetric embedded lens as
\bea
cT(\theta_S,\theta_I)&=& (1+z_d)\frac{D_dD_s}{D_{ds}}\Bigg[\frac{(\theta_S-\theta_I)^2}{2}
 +\theta_E^2\int_{x}^{1}\frac{f(x',z_d)-f_{\rm RW}(x')}{x'}dx'\Bigg].
\label{T}
\eea
It is straightforward to check that variation of Eq.~(\ref{T}) with respect to the image angle $\theta_I$ gives the lens  Equation~(\ref{spherical}).
Equation~(\ref{T}) is essentially the same as the Fermat potential for a spherical lens under the thin lens approximation in the standard linearized lens theory (Schneider et al.\ 1992) except for the $f_{\rm RW}$ term which accounts for the (lowest order) effects of embedding \citep{Kantowski13}.
The term ``Fermat potential" can refer to the two terms within the square brackets with or without the angular distances factor and/or the $1+z_d$ factor.
All factors are required for the use we make of it in this paper.
The geometrical and potential parts of the time delay, i.e., $T_g$ and $T_p,$ are respectively the first and the second terms in Eq.~(\ref{T}).
The key theoretical result in this paper is the relation of the {\it secondary} CMB temperature anisotropy \citep{Sachs67, Rees68} to a derivative of the potential part of the embedded lensing time delay,
\be
\frac{\Delta {\cal T}}{{\cal T}}= H_d\,\frac{\partial\, T_p}{\partial\, z_d},
\label{calT2}
\ee
where $H_d$ is the Hubble parameter at the deflector's redshift $z_d,$ and from Eq.\,(\ref{T})
\be\label{Tp}
cT_p(\theta_I,z_d)= 2(1+z_d)r_s\int_x^1{\frac{f(x',z_d)-f_{\rm RW}(x')}{x'}{dx'}}.
\ee
$T_p$ in Eq.\,(\ref{calT2}) can be replaced by the full time delay $T$ (hence the Fermat potential) because the geometrical part $T_g$ does not affect the CMB temperature, i.e., $\partial T_g/\partial z_d=0$ (the source angle $\theta_S$ as well as the image angle $\theta_I$ are held fixed during the differentiation).
We first investigate consequences of Eq.\,(\ref{calT2}) and postpone its subtle derivation to  Appendix~\ref{sec:appendix}.
A check of Eq.\,(\ref{calT2}) for an embedded static lens appears in Appendix~\ref{append:static}.

Once $T_p(\theta_I,z_d)$ is known for a given lens, e.g., via integrating Eq.~(\ref{Tp}), temperature fluctuations $\Delta {\cal T}$ in the CMB caused by that lens can easily be found by taking the  single derivative in Eq.~(\ref{calT2}).
For an arbitrary lens we can understand the source of $\Delta {\cal T}$ by differentiating Eq.~(\ref{Tp}) without performing the integration.
Because $z_d$ appears in two places in Eq.~(\ref{Tp}), two terms contribute to the derivative, i.e.,
$\Delta {\cal T}= \Delta {\cal T}_{\rm T}+\Delta {\cal T}_{\cal E},$
where one term is proportional to the potential part of the time delay and is called the time-delay part
\be
\frac{\Delta{\cal T}_{\rm T}}{{\cal T}}= \frac{H_d T_p}{1+z_d},
\label{calTT}
\ee
and a second term called the evolutionary part
\be
\frac{\Delta {\cal T}_{\cal E}}{{\cal T}}=(1+z_d)\frac{2r_s H_d}{c}\int_x^1\frac{dx'}{x'}\ \frac{\partial f(x^\prime,z_d)}{\partial z_d}.
\label{calTE}
\ee
Time delays, $T_p$ values in Eq.\,(\ref{calTT}), produced by individual galaxies are too small to produce detectable effects on the CMB
but time delays produced by large superclusters and huge voids are, see Fig.\,\ref{fig:dT} and Table \ref{tab:void}.
The evolutionary part is only present when the perturbation evolves at a different rate than the background cosmology.
We can relate these two parts to terms that appear when making a perturbative calculation of the ISW effect.
If $\chi$ is the comoving radial coordinate measured from the center of the embedded lens, if $\tilde{\chi}\equiv \chi/\chi_b$ is the normalized comoving distance, and if $\rho(\tilde{\chi}, z_d)$ and $\bar{\rho}(z_d)$ are the matter densities of the embedded spherical lens and the homogeneous FLRW background universe, then the comoving density contrast of the lens is $\delta(\tilde{\chi},z_d)\equiv(\rho-\bar{\rho})/\bar{\rho}.$
Using Eq.\,(\ref{Tp}) and the relation between projected and volume mass densities, the projected mass fraction can be written as
\be\label{fdelta}
f(x,z_d)-f_{\rm RW}(x)=3 \int_0^x dy\, y\int_0^{\sqrt{1-y^2}}d\zeta\,\delta(\sqrt{y^2+\zeta^2},z_d),
\ee
and the total temperature fluctuation can be written in terms of the lens density contrast $\delta$ as
\bea\label{Total}
\frac{\Delta{\cal T}}{{\cal T}}&=&\frac{6r_s}{c/H_d}\int_x^1\frac{dx'}{x'}\int_0^{x'} dy\, y\int_0^{\sqrt{1-y^2}} d\zeta\,
 \delta(\sqrt{y^2+\zeta^2},R_d)\left[1- \frac{R_d}{\delta}\frac{\partial\, \delta}{\ \partial R_d} \right].
\eea
The first and second terms within the square brackets correspond to the time-delay and evolutionary contribution, $\Delta {\cal T}_T$ and $\Delta {\cal T}_{\cal E}$, respectively.
In Eq.\,(\ref{Total}), $R_d\equiv R(t_0)/(1+z_d)$ and is used as a `time' variable rather than $t_d$.
There is no time delay or temperature fluctuation at the boundary of the Swiss cheese void where $x=1$ (i.e., $\theta_I=\theta_M$) for any embedded lens.
For linearly evolving structures in an Einstein-de Sitter (EdS) universe, $\delta\propto R_d$  and the two terms in Eq.\,(\ref{Total}) exactly cancel resulting in $\Delta{\cal T}= 0,$ consistent with the classical results of \cite{Sachs67}.
For linearly evolving structures in the standard $\Lambda$CDM universe, $|\delta|$ grows slower than $R_d,$ and the evolutionary contribution only partially cancels the time-delay contribution resulting in a non-zero ISW effect (see Fig.~\ref{fig:ISW} for an example).
For highly non-linear structures \citep{Bertschinger85a, Bertschinger85b, Sheth04}, $|\delta|$ might cease to grow or even start to decrease (e.g., a deep cosmic void with density contrast $\delta$ already approaching the lower bound -1), the evolutionary contribution can be of the same sign as the time-delay contribution.
For such cases the time-delay contribution $\Delta {\cal T}_T$ can be used as a conservative estimate of the magnitude of the ISW signal (see the examples in the next Section).
For embedded point mass lenses or lenses that expand with the background $f(x,z_d)=f(x)$ does not explicitly depend on $z_d$ and the temperature fluctuation is simply proportional to the time delay, see  Eq.\,(\ref{calTT}).
The time-delay contribution to the temperature fluctuation is of the order $H_dT_p\sim H_d(2 r_s/c)\sim 2\beta_d r_s/r_d$
where $c\beta_d\equiv H_d r_d$ is the expansion rate of the Swiss cheese void's boundary relative to static observers and $r_d$ is the physical radius of the void at redshift $z_d$ \citep{Kantowski13}.
This is in line with estimates originally made in Rees \& Sciama (1968) although the details differ.
From Eq.\,(\ref{calTE}) the evolutionary contribution is expected to be the same order of magnitude as long as only moderate expansion/contraction occurs.


\section{Examples}\label{sec:examples}


\begin{figure*}
\begin{center}$
\begin{array}{cc}
\hspace{-15pt}
\includegraphics[width=0.53\textwidth,height=0.35\textheight]{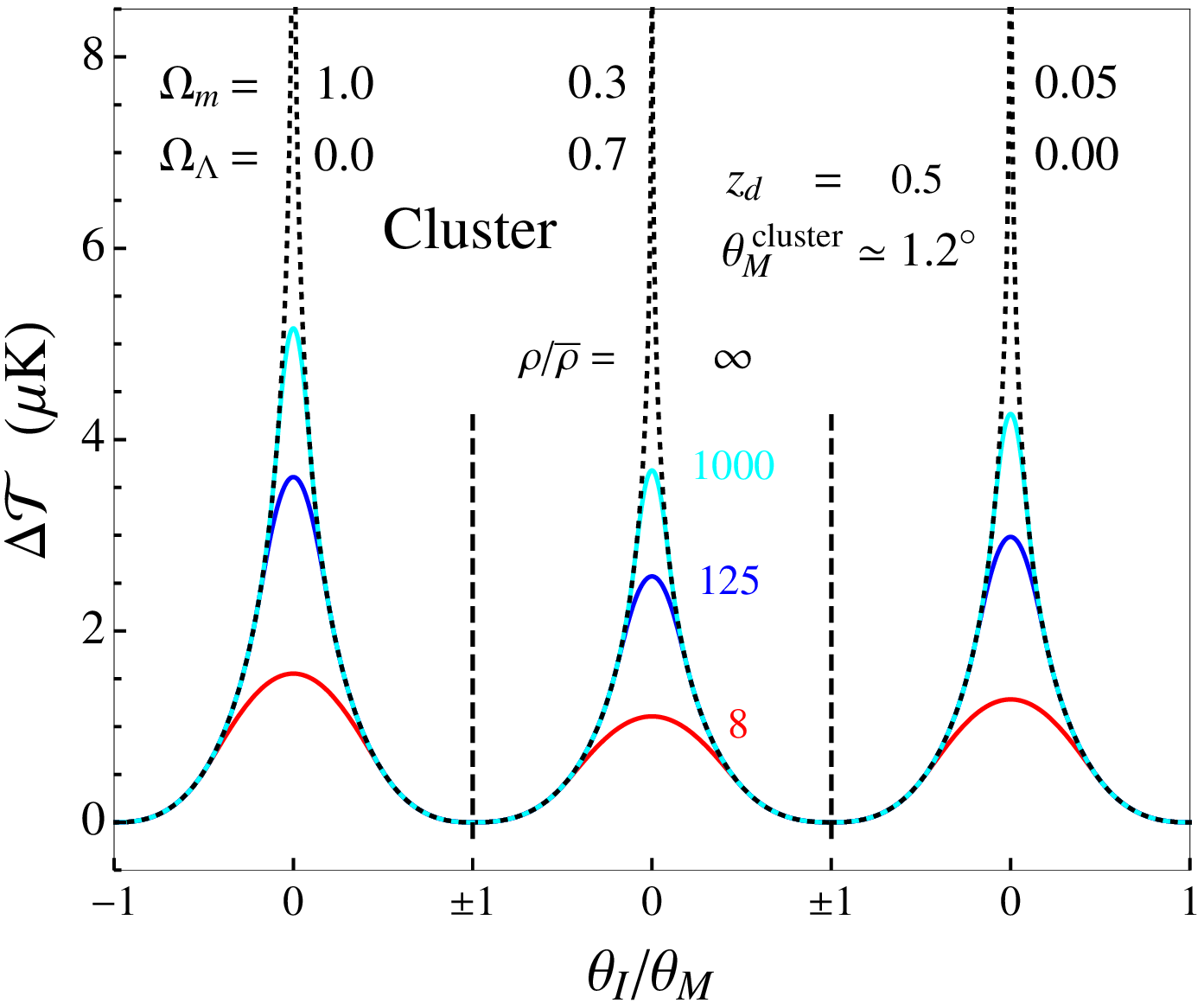}
\hspace{5pt}
\includegraphics[width=0.53\textwidth,height=0.35\textheight]{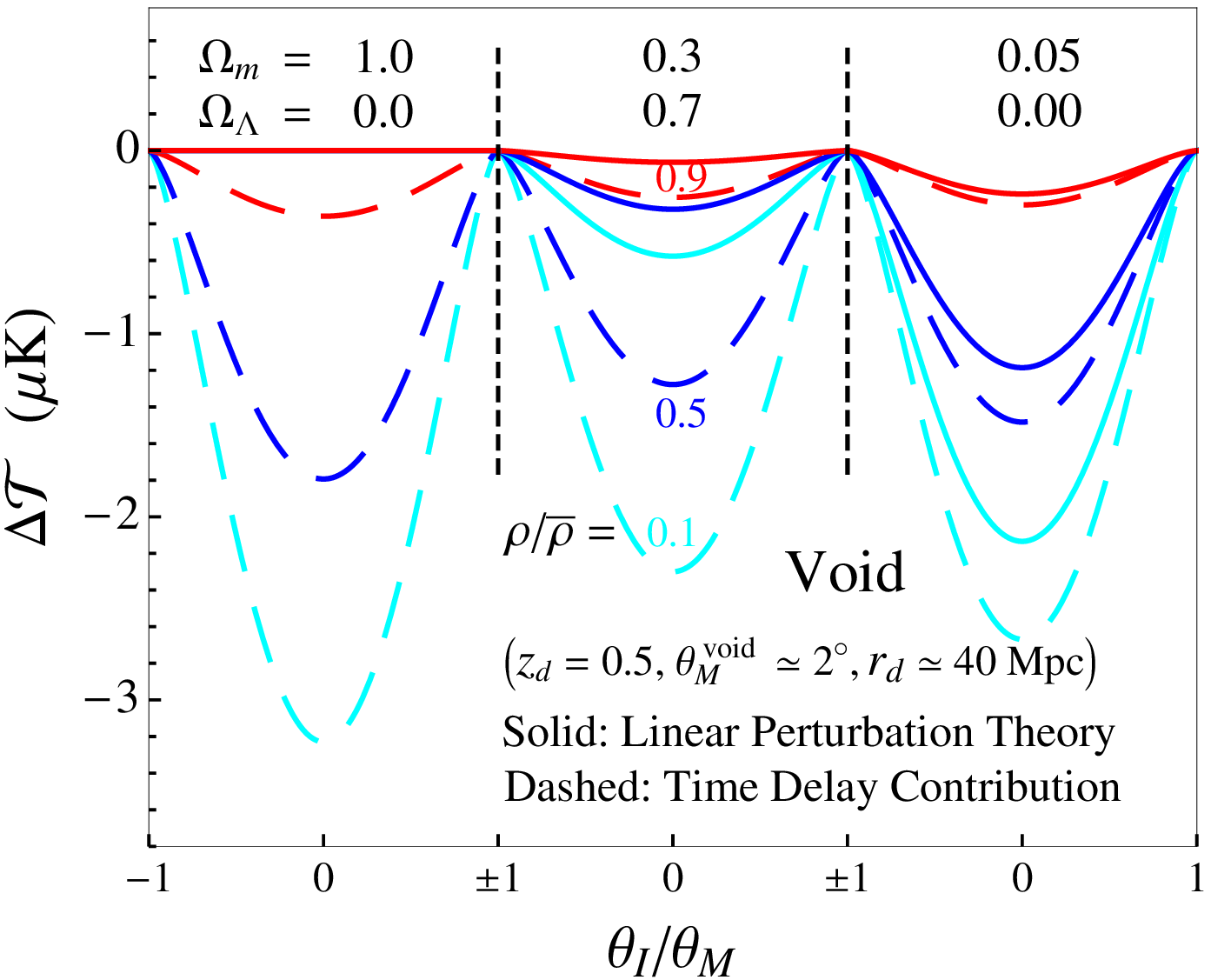}
\end{array}$
\end{center}
\caption{ Secondary anisotropies in the CMB across an embedded super cluster lens of angular size $\theta_M\simeq 1.2^\circ$ and a void lens of size $\theta_M\simeq2^\circ$ (respectively left and right panels) both at $z_d=0.5$.
Three background FLRW cosmologies are shown in each panel: the Einstein-de Sitter universe (left curves), the concordance $\Lambda$CDM universe (middle curves), and an open baryonic-matter only universe (right curves).
The concordance cluster and void lens masses are about $10^{16}M_{\odot}$ and $4.8\times 10^{16}M_{\odot},$ respectively.
For the cluster lens  $\Delta {\cal T}$ (in $\mu$K) is shown for four models: three uniformly distributed mass condensations with central over-densities $\rho/\bar{\rho}=8,$ 125, and 1000, all co-expanding with the background cosmologies (the respective solid red, blue and cyan curves),  and for a point mass condensation (the dotted curves).
In the right panel cosmic voids are modeled as uniformly under-dense spherical regions of densities $0.1\bar{\rho},$ $0.5\bar{\rho},$ and $0.9\bar{\rho}$ bounded by thin walls containing the remaining mass of the removed homogeneous sphere (the cyan, blue, and red curves, respectively).
We assume the void to be either co-expanding, the dashed curves ($\Delta{\cal T}$ is produced entirely by the time-delay contribution of Eq.\,(\ref{calTTvoid})) or evolving according to linear perturbation theory, the solid curves ($\Delta{\cal T}$ is produced by the sum of Eqs.\,(\ref{calTTvoid}) and (\ref{calTEvoid}) with appropriate values for $d\log\delta/dz_d$, see Fig.~\ref{fig:ISW}).}
\label{fig:dT}
\end{figure*}


\begin{figure*}
\begin{center}
\includegraphics[width=0.9\textwidth,height=0.45\textheight]{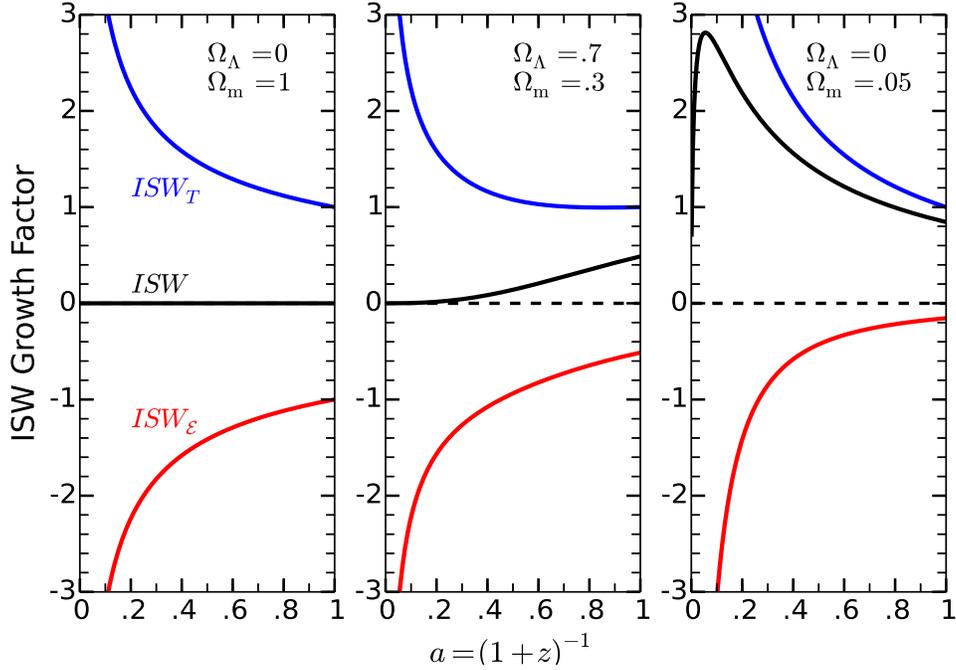}
\end{center}
\caption{ Redshift dependence of the ISW effect for the void models described in Section\,\ref{sec:examples}.
We have tested three background FLRW cosmologies: EdS, concordance $\Lambda$CDM, and an open baryonic-matter only universe.
The three curves in each panel correspond to the time delay contribution $\Delta {\cal T}_T(z)$, evolutionary contribution $\Delta {\cal T}_{\cal E}(z),$ and the sum $\Delta {\cal T}=\Delta {\cal T}_{T}+\Delta {\cal T}_{\cal E}$ (the blue, red, and black curves, respectively).
All curves are normalized by $\Delta {\cal T}_T(z=0)$ for the corresponding cosmology.
We have assumed a linear perturbation growth of voids when computing $\Delta {\cal T}_{\cal E}(z)$ and $\Delta {\cal T}(z),$ see Eqs.~(\ref{calTTvoid}) and (\ref{calTEvoid}).
}
\label{fig:ISW}
\end{figure*}

\begin{figure*}
\begin{center}$
\begin{array}{cc}
\hspace{-15pt}
\includegraphics[width=0.52\textwidth,height=0.35\textheight]{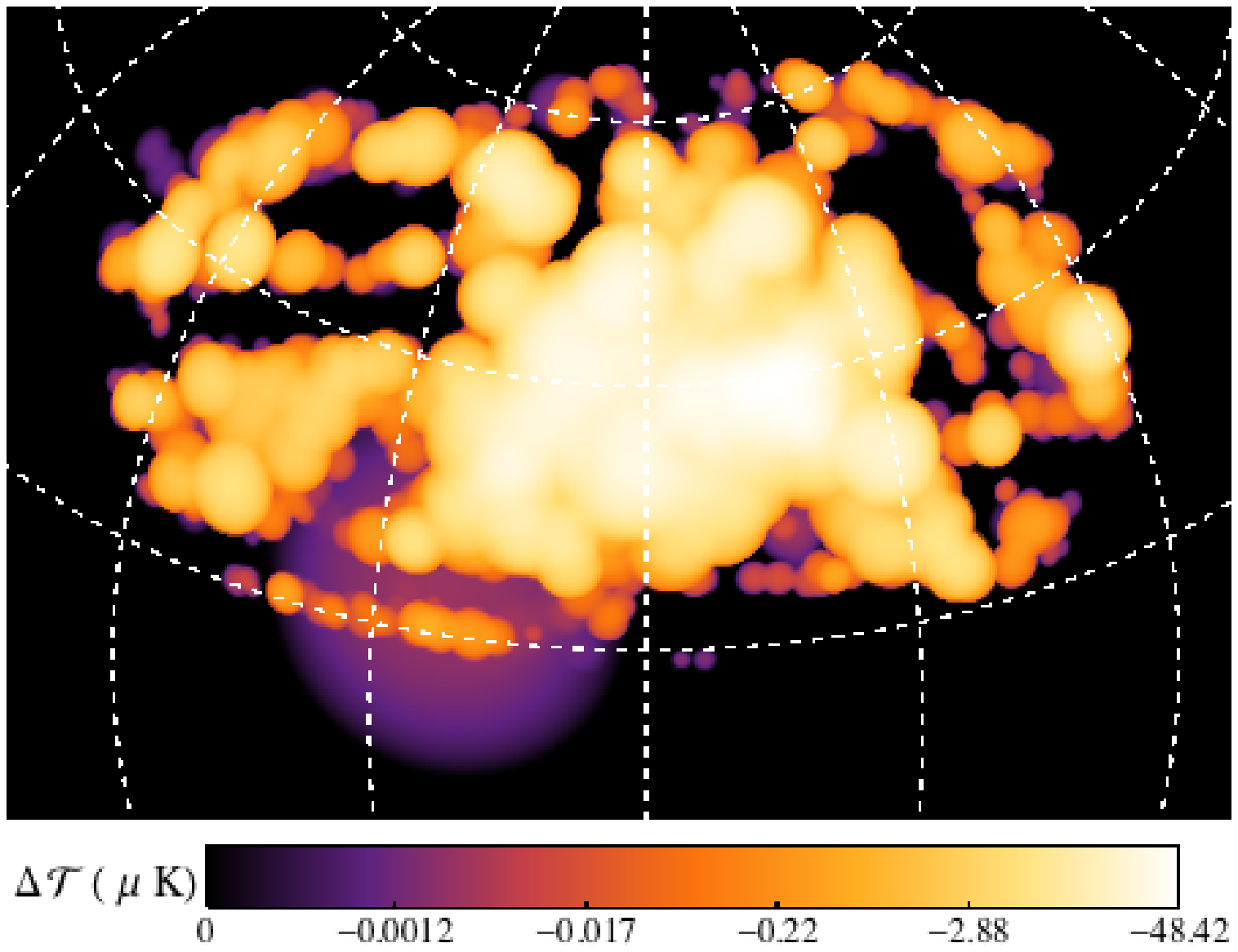}
\hspace{10pt}
\includegraphics[width=0.52\textwidth,height=0.35\textheight]{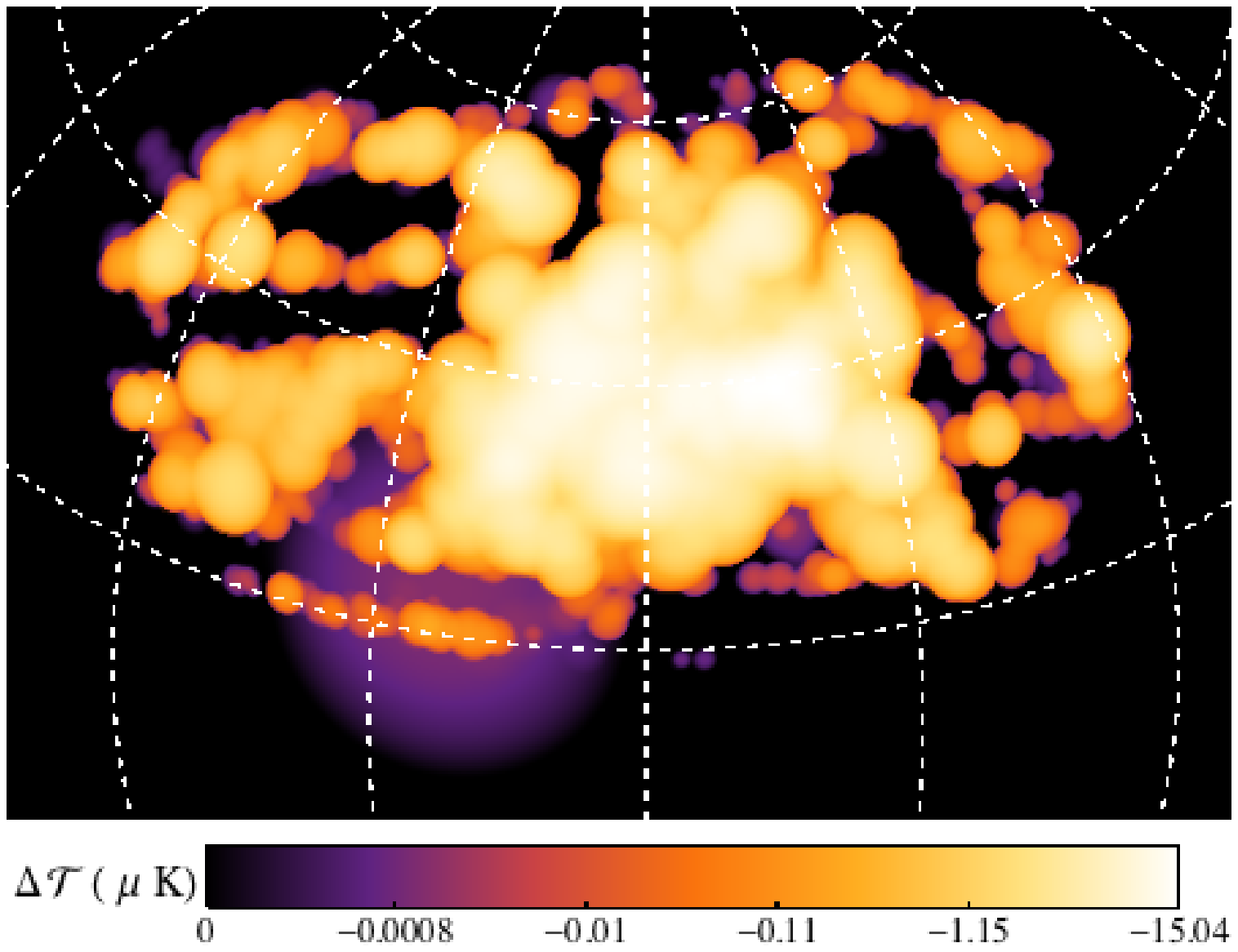}
\end{array}$
\end{center}
\caption{The Lambert equal-area projection of the sky map of CMB temperature fluctuations caused by  passing through voids along the line of sight.
        We used the ``central void catalog" (version 2014.06.18) of Sutter et al.\ (2012).
        The map is centered at $\rm RA=180^\circ$ and $\rm DEC=30^\circ$ with longitude and latitude lines separated by $30^\circ.$
        We use the same void model as in the right panel of Fig.~\ref{fig:dT} (the blue curves in the middle panel).
        For each void, we estimate its mass (or Schwarzschild radius $r_s$) using its effective radius and redshift provided by \cite{Sutter12}, we then compute $\Delta{\cal T}$ across this void using Eq.\,(\ref{calTTvoid}) assuming $\delta=-$0.5.
        For overlapping voids along a sight line, we add the $\Delta{\cal T}$ caused by each void.
        For the left panel, we assume the voids are co-expanding with the background Universe and only the time delay contributes.
        For the right panel, we assume a linear perturbation theory (see Fig.\,\ref{fig:ISW}).
        Assuming a different void density contrast will also result in a similar figure except for a pure rescaling of the color-bar.
        For example, if we chose $\delta =-0.1$ instead of $\delta=-0.5,$ then the signal will be reduced to  one fifth of its current values for both scenarios.
        }
\label{fig:Sutter}
\end{figure*}

To illustrate the simplicity of the embedded theory's use, as well as to give an order of magnitude estimate of the effect on the CMB's temperature fluctuations, we first construct simple embedded lens models for galaxy clusters and cosmic voids and then compute the late-time ISW effect by applying Eq.~(\ref{calT2}) to these models.
We consider three centrally over-dense lenses (galaxy clusters) and three centrally under-dense lenses (cosmic voids).
We assume the central mass densities of the embedded cluster lenses are homogeneous and expanding with the background, i.e., have constant angular radii $\overline{\theta}_M<\theta_M$ (see  the central shaded region of Fig.~\ref{fig:cheese}).
The projected mass fraction for this lens is $f_{\rm cluster}(x)=f_{\rm RW}(x/\bar{x})$ for $x\le\bar{x}$ and equals 1 for $\bar{x}<x\le1,$ where  $\bar{x}\equiv \bar{\theta}_M/\theta_M.$
From Eqs.~(\ref{Tp}) and (\ref{calTT}) we obtain the temperature anisotropy for this simple cluster lens model
\be\label{dTcluster}
\frac{\Delta {\cal T}}{\cal T}=\frac{H_d}{1+z_d}\times\begin{cases}
      & { T}_p^{{\rm PM}}(x)-{ T}_p^{{\rm PM}}(x/\bar{x}),  \text{  if $x\le\bar{x}$ }, \\
      & { T}_p^{{\rm PM}}(x), 					  \text{\hspace{65pt} if   $\bar{x}<x\le 1$},
\end{cases}
\ee where ${ T}_p^{{\rm PM}}(x)$ is the Point Mass time delay given in  Eq.~(7) of \cite{Kantowski13}
\be
{ T}_p^{{\rm PM}}(x) =  (1+z_d)\frac{2r_s}{c}\left(\log\left[\frac{1+\sqrt{1-x^2}}{x}\right] -\frac{4-x^2}{3}\sqrt{1-x^2}\right).
\label{TPM}
\ee
In the left panel of Fig.\,\ref{fig:dT}
we give $\Delta {\cal T}$ for  $\xo=  0.5,$ $0.2$, and $0.1$, corresponding to respective density contrasts $\rho/\bar{\rho}$ of 8, 125, and 1000.

For voids the lensing picture is similar to Fig.\,\ref{fig:cheese} except the central region is an under-density and the deflection  angle $\alpha$ is outward, i.e, a void is de-focusing rather than focusing.
To be an exact solution of GR, Swiss cheese embedding requires the under-dense void interior to be surrounded by an over-dense bounding ridge. 
For simplicity we model a cosmic void as a uniform under-dense spherical region of density $\rho/\bar{\rho}=1+\delta=0.1, 0.5,$ or 0.9 (i.e., having density contrasts $\delta=-0.9$, $-0.5$ or $-0.1$, see the right panel of Fig.\,\ref{fig:dT}) bounded by an over-dense thin shell which contains the mass removed from the homogeneous sphere (i.e., $f_{\rm shell}=-\delta\,[1-(1-x^2)^{1/2}$]). 
If the voids are co-expanding $\delta$ remains constant; however, if they evolve $\delta$ changes with the deflector's redshift $z_d$.
For this void model
\be
\frac{\Delta{\cal T_T}}{\cal T}=\delta\frac{2r_s}{3c/H_d}(1-x^2)^{3/2},
\label{calTTvoid}
\ee
and
\be
\frac{\Delta{\cal T_{\cal E}}}{\cal T}=(1+z_d)\frac{d\log\delta}{d\,z_d}\times\frac{\Delta{\cal T_T}}{\cal T}.
\label{calTEvoid}
\ee
For simplicity we consider only two possibilities: the co-expanding case  where $\delta'(z_d)\equiv d\delta/dz_d\equiv 0$ and only $\Delta {\cal T}_T$ contributes, and a linear evolving case where $\delta(z_d)$ is known from standard perturbation theory.
Despite the extreme simplicity of this result, the model is useful for studying lensing by voids \citep{Chen13}.
Extension to other cluster or void models is straightforward.
For example if the compensating ring of mass at the boundary of the Swiss cheese void in this simple model is moved
inward it can produce a hot ring surrounding the void's central cold spot.

\begin{deluxetable}{lcccclll}
\tabletypesize{\footnotesize}
\tablecolumns{8}
\tablewidth{0pt}
\tablecaption{ Amplitudes of ISW cold spots produced by compensated voids in the concordance $\Lambda$CDM cosmology.
CMB temperature reductions in $\mu$K are given for Swiss cheese voids of radii $r_d= $ 5, 20, 50, and 100 Mpc at redshifts $z= 0.5$ and $1.0$ with density contrasts $\delta=(\rho-\overline{\rho})/\overline{\rho}= -0.1,-0.5,$ and  $-0.9$.
The comoving cosmic mass $M_d$ contained in the void radius $r_d$ is also given as well as its angular size $\theta_M$.
\label{tab:void}}
\tablehead{
\colhead{$\Delta {\cal T} \>\>(\rm\mu\, K)$}&
\colhead{$r_d$ = 5 Mpc }&
\colhead{20 Mpc }&
\colhead{50 Mpc  }&
\colhead{100 Mpc }
}
\startdata
$z_d=0.5$  &  $M_d=\, 7.2\times 10^{13}$ M$_\odot$  & $4.6\times 10^{15}$ M$_\odot$ 	&  $7.2\times 10^{16} $M$_\odot$    & $5.8\times 10^{17} $M$_\odot$ 	  \\
 &  $\theta_M =\, 0.23^\circ$ & $0.91^\circ$ &  $2.3^\circ$   & $4.6^\circ$ 	  \\
$ \delta=-0.1$  &  $(-0.0001, -0.00038)$\tablenotemark{a}  & $(-0.006, -0.025)$	& $(-0.096,-0.38)$  		& $(-0.77,-3.06 )$	  \\
$ \delta=-0.5$  &  $(-0.0005, -0.0019)  $  & $(-0.031, -0.12)$	& $(-0.479,-1.91)$ 		& $(-3.83,-15.29)$	  \\
$ \delta=-0.9$  &  $(-0.0009, -0.0034)  $  & $(-0.055, -0.22)$	& $(-0.861,-3.44)$ 		&  $(-6.89,-27.52)$ 	  \\
\hline
\\

$z_d=1.0$  &  $M_d=\, 1.7\times 10^{14}$ M$_\odot$  & $1.1\times 10^{16}$ M$_\odot$ 	&  $1.7\times 10^{17} $M$_\odot$    & $1.4\times 10^{18} $M$_\odot$ 	  \\
 &  $\theta_M =\, 0.17^\circ$ & $0.69^\circ$ &  $1.7^\circ$   & $3.5^\circ$ 	  \\
$\delta=-0.1$  	&  $(-0.00016,-0.0012)  $  & $(-0.01,-0.078)$	& $(-0.16,-1.22)$  		& $(-1.27,-9.77)$ 	  \\
$ \delta=-0.5$  	&  $(-0.0008,-0.0061)  $  	& $(-0.05,-0.39)$	& $(-0.79,-6.10)$ 		& $(-6.35,-48.85)$	  \\
$\delta=-0.9$  	&  $(-0.0014,-0.011)    $  	& $(-0.09,-0.70)$	& $(-1.43,-10.97)$ 		&  $(-11.43,-87.93)$ 	
\enddata
\tablenotetext{a}{ The number pairs are for linearly evolving voids (left) and co-expanding voids (right) respectively.
The value for the linearly evolving case is about a factor of 0.25 and 0.13 of the value for the co-expanding case for $z_d=0.5$ and $1.0$, respectively (see Fig.~\ref{fig:ISW}).
}
\end{deluxetable}

In the left panel of Fig.~\ref{fig:dT} we plot the radial profile of the ISW effect across a super cluster at redshift $z_d=0.5$ with an angular radius $\theta_M^{\rm cluster}\simeq 1.2^\circ.$
The cluster lens mass is $\sim$$10^{16}M_{\odot}.$
We test for three background cosmologies: an EdS universe ($\Omega_{\rm m}= 1.0,$ $\Omega_{\Lambda}= 0$), a concordance $\Lambda$CDM universe ($\Omega_{\rm m}= 0.3,$ $\Omega_{\Lambda}= 0.7$), and a baryonic-matter only universe ($\Omega_{\rm m}= 0.05,$ $\Omega_{\Lambda}= 0$), i.e., the left, middle, and right curves, respectively.
From the left panel of Fig.~\ref{fig:dT} we see that for a cluster lens of fixed mass, the amplitude of the perturbation increases with central concentration (approaching the limiting case of an embedded point mass lens; $f_{\rm point}=1$).
The CMB temperature increases by $\gtrsim 2\,\mu$K for a co-expanding cluster of $\sim$$10^{16}M_\odot$ with density contrast $\delta\gtrsim100.$
The ISW signal predicted by the simple cluster model scales with the cluster mass, see Eqs.~(\ref{dTcluster}) and (\ref{TPM}).
Outside of the over-dense region, i.e., the vacuum, the cluster behaves exactly the same as a (static) point mass lens;
inside the over-dense region, the lens might be still evolving. 
Consequently, prediction based on the simple model might be biased (e.g., increases more slowly near the cluster center than predicted in Fig.~\ref{fig:dT}).
However, since the over-dense region takes only a small fraction of the lens volume, our model should give reasonable estimate for the averaged signal for rich galaxy clusters ($\langle\Delta T\rangle $ about $\sim$$0.3\,\mu$K for a  $10^{16}M_{\odot}$ cluster assuming $\delta\gtrsim100$).
More physical embedded cluster model (e.g., non-uniform density profile with different evolution scenarios) will be pursued in future works (e.g., Kantowski et al.\ 2014).
We also see that various background cosmologies influence $\Delta {\cal T}_{T}$ through the Hubble parameter $H_d$; however, the differences are not very significant.

The right panel of Fig.~\ref{fig:dT} and Table~\ref{tab:void} show predictions for the two simple void lens models (co-expanding and linearly evolving) at redshift $z_d=0.5.$
The concordance void lens has angular radius $\theta_M^{\rm void}\simeq2^\circ$ with physical radius $r_d\approx 40\,\rm Mpc$ and mass $\sim$$4.8\times 10^{16}M_{\odot}.$
The dashed and solid curves show the results for co-expanding and linearly evolving void  models, respectively.
The signal scales linearly with the lens mass $r_s$ (equivalently as the cube of the void size $\theta_M$) and emptier voids result in colder CMB spots, i.e., they scale with $\delta$.
From Fig.~\ref{fig:dT} the CMB temperature is seen to decrease by $\gtrsim1.5\,\mu$K for a co-expanding void of radius $\theta_M\gtrsim2^\circ$ if $\delta\lesssim-0.5$.
For a void evolving relative to the background cosmology there is an additional  contribution to the temperature profile, the evolutionary part.
If the void evolves according to standard linear perturbation theory, then the evolutional contribution $\Delta T_{\cal E}$ of Eq.\,(\ref{calTEvoid}) significantly cancels the time-delay contribution $\Delta {\cal T}_{T}$ of Eq.\,(\ref{calTTvoid})  (see Fig.\,\ref{fig:ISW}).
The cancellation is complete  for an EdS cosmology as expected, and the total $\Delta {\cal T}$ is reduced by 75\% and 20\% of the time-delay contribution respectively for $\Lambda$CDM and baryonic-matter only universes.
Cosmic voids discovered from galaxy surveys have radii from $\sim$$5\,\rm Mpc$ to $\sim$$100\,\rm Mpc$ with majority of them having radii of the order of $\sim$$10\,\rm Mpc$ (Sutter et al.\ 2012).
In Table~\ref{tab:void} we estimate the magnitude of the ISW cold spot caused by cosmic voids having different void sizes and density contrasts.
We present results for both the linearly evolving and co-expanding scenarios and test for two redshift $z_d=0.5$ and $1.$
For small density contrast (e.g., $\delta=-0.1$) linear perturbation theory should give a better estimate of the ISW signal, whereas for large density contrast (e.g., $\delta=-0.9$), the co-expanding model should give a better estimate.
The two numbers can be used as  approximate lower and upper bounds for the ISW signal caused by cosmic voids.

Figure~\ref{fig:ISW} shows the redshift dependence of the amplitude of the ISW effect for the two simple void lens models presented in this paper.
We plot the growth curves of the ISW effect with redshift for the time-delay contribution $\Delta {\cal T}_{T}$ , evolutionary contribution $\Delta{\cal T}_{\cal E}$ assuming linear perturbation theory,
and the sum of the two contributions (the blue, red, and black curves respectively).
To compute the ISW temperature profiles for a void at an arbitrary redshift $z$, you simply scale the curves in the right panel of Fig.~\ref{fig:dT} ($z_d=0.5$) using the corresponding curves in Fig.~\ref{fig:ISW}, i.e., apply a multiplicative factor $\rm ISW_T(z)/ISW_T(0.5)$  and $\rm ISW(z)/ISW(0.5)$ for the co-expanding and linear evolving model respectively.
The reader can observe from the left panel, as stated earlier, that linearly evolving perturbations in an EdS universe do not directly alter the CMB's temperature, i.e., the net growth factor vanishes ${\rm ISW}(z)=0$ for all $z$.
What is clear for the other two cosmologies is that for the highest redshifts ($z>20$) the earlier in cosmic time they are encountered the less they perturb the CMB.
However, somewhat closer voids ($20>z>0$) in an open baryonic matter universe (see the black curve in the third panel) will produce  larger effects on CMB temperatures with increasing redshifts, whereas, voids in a concordance universe (see the black curve in the middle panel) have a rapidly diminishing effect (with increasing redshift) on the CMB for all $z>0$.
Seeing a decrease in ISW temperatures with redshift is clearly expected if we do live in a concordance universe.

Figure~\ref{fig:Sutter} shows an example of ISW-induced CMB temperature fluctuations (with and without linear evolution) using the recent void catalogue from SDSS DR7 \citep{Sutter12,Abazajian09} and the void models described above.
For simplicity we have assumed the same density contrast $\delta=-0.5$ for all voids (density profiles for individual comic voids are not available, only mean profiles have been estimated by stacking/averaging a large number of cosmic voids).
The majority of the voids have radii of the order of $\sim$$10\, \rm Mpc$, and their evolution can be non-linear.
Consequently, the time-delay contribution $\Delta {\cal T}_{T}$ (the left panel) might give better estimates of the ISW signals.
For those really large voids (with radius $\sim$$\rm 80\, Mpc$), if they are shallow as predicted by standard theory, then linear perturbation theory should give better estimates of the signal (the right panel).
Because most of the voids are at low redshifts ($z\lesssim0.5$), the ISW signal of the linear evolution scenario is about one third to one fourth of the time-delay contribution because of the cancellation of $\Delta {\cal T}_{T}$ by $\Delta {\cal T}_{\cal E}$ (see Fig.~\ref{fig:ISW}).
Consequently, the right panel looks very similar to the left panel except for the color-bar.
Assuming a different void density contrast will also result in a similar figure except for a pure rescaling of the color-bar.
For example, if we chose $\delta =-0.1$ instead of $\delta=-0.5,$ then the signal will be reduced to one fifth of its current values for both scenarios.

\section{Discussions and Conclusions}\label{sec:conclusions}

\citet{Granett08a} claimed a detection of the ISW effect at $\rm -11.3\pm 3.1\mu K$ by stacking 50 of the largest voids (effective radius $\gtrsim 100\,\rm Mpc$) from their catalog using the WMAP 5-year data.
Similar analyses have also been conducted for the same and other void catalogs (Sutter et al.\ 2012; Pan et al.\ 2012; Nadathur \& Hotchkiss 2014) by several other groups using WMAP or \emph{Planck} data (e.g., Ili\'c et al.\ 2013; Cai et al.\ 2014; Planck Collaboration 2014b; Hotchkiss et al.\ 2015).
Analyses of the more recent catalogs  have typically given signals of more moderate amplitude and lower significance (from negligible to $\sim$$2.5\,\sigma$; Ili\'c et al.\ 2013; Planck et al.\ 2014b).
For example, Ili\'c et al.\ (2013) confirmed the Granett et al.\ (2008a) results (at slightly lower significance) using WMAP 7-year data.
They detected no signals using the Pan et al.\ (2012) catalog and a signal up to $\sim$$2.4\,\sigma$ using the Sutter et al. (2012) catalog (after rescaling of the voids to the same size on the stacked images).
Recently the amplitude of the Granett et al.\ (2008a) detection has been re-confirmed by \emph{Planck} by applying the aperture photometry method to \emph{Planck} CMB temperature maps and the same void catalog (Planck Collaboration 2014b).
Several groups found that the observed CMB fluctuations from stacking and averaging cosmic voids are significantly larger than expected from the linear ISW effect.
For example, \citet{Nadathur12} found a theoretical value from a linear perturbation analysis in $\Lambda$CDM cosmology $\rm \gtrsim -2\mu K,$ inconsistent with the observed value from \citet{Granett08a} at $>3\,\sigma.$
Numerical simulations have also predicted smaller effects \citep{Maturi07, Cai10, Cai14, Flender13, Watson14, Hotchkiss15}.
For example, Cai et al.\ (2014)  found a cold ISW imprint of amplitude between 2.6 and 2.9 $\mu K$ and significance of $\sim$$2\,\sigma$ using voids selected in the SDSS DR7 spectroscopic redshift galaxy catalog calibrated by  N-body simulations.
A $\Lambda$CDM signal consistent with zero was predicted  by Hotchkiss et al.\ (2015) using the Jubilee N-body simulation and confirmed using a catalogue of voids and superclusters extracted from  SDSS, thus challenging all positive ISW interpretations. 

We tentatively compare predictions based on our simple embedded void model with recent observations based on void-stacking (Granett et al.\ 2008a; Planck Collaboration 2014b).
The time delay of a deep cosmic void ($\delta=-0.9$) of angular radius $\sim$$4^\circ$ at $z_d=0.5$ (about the mean radius and redshift for the 50 voids in Granett et al.\ 2008a) results in $\Delta {\cal T}\approx -18.8\,\mu K$ toward the void center and an averaged cold spot of about $-7.5\,\mu K$ after convolving with a compensated top-hat filter of aperture radius $\theta_F= 4^\circ.$
These temperature reductions are of the same order as observed from stacking and averaging large cosmic voids in the Granett et al.\ (2008a) catalog \citep{Granett08a,Planck14b}.
However, to produce an effect of this order of magnitude requires those large voids to be very deep,  $\delta\sim-0.9$ and/or possibly evolve in an unknown non-linear way.
Cosmic voids in galaxy surveys were selected as regions with low galaxy number densities; however, luminous matter as a tracer of dark matter might be biased.
The dark matter density contrast within a cosmic void might be shallower than the $\delta\sim-0.9$ estimate based on galaxy number counts \citep{Sutter12, Sutter14}.
If the void is much shallower, as predicted by standard theory, then the ISW effect will not be able to produce signals of the order of the recent detections.
For example, if we assume $\delta=-0.5$ (still a large density contrast), then estimates based on time delays alone are not sufficient because the void is probably still evolving and the evolutionary contribution needs to be included (see Fig.~\ref{fig:ISW}).
For linear evolution, our model predicts a signal of the order of $\sim$$2\,\rm \mu K$, similar to results of other predictions \citep{Nadathur12,Flender13,Ilic13,Cai10,Cai14}.
Calibrating the galaxy bias for void catalogs is an active field of research (Rasset et al.\ 2007; Szapudi et al.\ 2014a,b; Hamaus et al.\ 2014; Leclercq et al.\ 2014).
A linear galaxy bias $b_g = 1.41\pm 0.07$ was recently used by Szapudi et al. (2014a,b) to model the CMB Cold Spot using the ISW effect caused by a super giant void of radius $192\pm15\, h^{-1}\,\rm Mpc$ and density contrast $\delta=-0.13\pm0.03$ (see also Finelli et al.\ 2014).
Similar estimate on $b_g$ was obtained earlier by Rassat et al. (2007) for galaxies selected from the Two Micron All Sky Survey (Jarrett et al.\ 2000).
The galaxy void profiles of Sutter et al. (2012) suggest very low density near the void center, $\delta\lesssim -0.8$ for voids of radii range from a few up to $\sim$$100\,\rm Mpc.$
For simplicity, we have assumed a uniform dark matter density contrast $\delta=-0.5$ for all voids (large and small) when simulating the CMB sky map modulated by ISW effect caused by cosmic voids (see Fig~\ref{fig:Sutter}).
This is equivalent to assuming a linear galaxy bias $b_g\approx 1.8$ for all voids (see Fig.~9 of Sutter et al.\ 2012).
For simplicity, we have also assumed all voids to be compensated (with zero net mass with respect to the homogeneous background).
Numerical simulations show that small voids tend to be over-compensated, while large voids might be under-compensated (Sheth \& van de Weygaert 2004; Ceccarelli et al.\ 2013; Hamaus et al.\ 2014; Cai et al.\ 2014). 
More accurate modeling using more physical density profiles and better estimates of galaxy bias (Sutter et al.\ 2014; Hamaus et al.\ 2014; Leclercq et al.\ 2014) is beyond the scope of the paper but is important to pursue in further work.

The new results of this paper are Eqs.\,(\ref{T})--(\ref{Tp}), which result in Eqs.\,(\ref{calTT})--(\ref{Total}), i.e., simple analytical expressions for  {\it secondary} temperature anisotropies in the CMB caused by encountering mass inhomogeneities, i.e., embedded gravitational lenses.
The temperature fluctuations are obtained from a simple derivative of the Fermat potential.
That derivative is the same as a derivative of the potential part of the time delay which is easy to compute for an arbitrary lens model via Eq.\,(\ref{Tp}).
This new method of obtaining CMB temperature changes represents a significant improvement over previous methods.
We find that the CMB temperature change, i.e., $\Delta{\cal T}$ given in Eq.\,(\ref{calT2}), can be understood as coming from two sources.
The first source of change is the time delay and occurs because the increase in the Swiss cheese void's radius depends on how long the lens mass delays the passing photon.
The longer it is delayed the larger the cosmic expansion and the more the unlensed CMB photons are redshifted.
This term is given in Eq.\,(\ref{calTT}).
Clusters tend to delay lensed photons making them bluer and voids shorten transit times making them redder.
A second source of temperature change is the evolution of the distribution of the embedded mass inside the Swiss cheese void relative to the background cosmology,
i.e., a $f$ and $\delta$ dependence on cosmic time alters the frequency of the transiting photon.
This temperature change can be positive or negative depending on lens details and is given in Eq.\,(\ref{calTE}).
We have confirmed Eqs.\,(\ref{calTT}) and (\ref{calTE}) by comparing results with prior analytic examples, e.g., the embedded expanding homogeneous dust sphere model  \citep{Dyer76}, and the compensated void/lump model under the potential approximation \citep{Martinez90}.
We have used Eq.\,(\ref{calTE}) in Fig.\,\ref{fig:ISW} to estimate the effect of linear perturbation growth of the voids temperature profiles of Fig.\,\ref{fig:dT}.
Our theory is valid for arbitrarily large density contrast and is therefore very useful for studying CMB lensing in nonlinear regimes, e.g., lensing by cosmic voids where the density contrast can be large \citep{Sutter12,Sutter14} and by galaxy clusters where $\delta\gg1$.
For individual nonlinear cosmic structures, the temperature fluctuation can be easily computed provided that the expanding/collapsing histories of the structures are known from astrophysical theories or simulations \citep{Press74,Bertschinger85a,Bertschinger85b,Sheth04}.
This is in contrast to the standard linear ISW analysis where the Poisson equation for the perturbation potential (or the density contrast) is solved in Fourier space with some assumed initial power spectrum.
Such an approach is not convenient for modeling temperature fluctuations caused by individual structures with a fixed scale, whereas our embedded theory is. 
We found that to produce secondary anisotropies of the order of some recent measurements of the ISW effect made using the void-stacking method requires large cosmic voids to be more numerous, deeper, and evolving differently  than predicted by linear perturbation theory.

\appendix

\section{ Proof of Equation\,(\ref{calT2})}\label{sec:appendix}

Suppose two identical CMB photons with frequencies $\nu_{\td}(t_1)$ reach the embedded lens at comoving point 1 in Fig.\,\ref{fig:cheese} at cosmic time $t_1$.
One photon is gravitationally deflected by the embedded lens  and exits at comoving point 2 at time $t_2$ on a path differing in direction by an angle $\alpha$.
The second photon is reflected by an angle $\alpha$ at point B as it travels from point 1 to point 2 on two straight line segments, just  as if the embedded lens didn't exist.
The lensed photon exits comoving point 2 at cosmic time $t_2$ which is later than the reflected photon by an amount of time $T_p(t_2)$ which is by definition the potential part of the time delay \citep{Chen10}.
After leaving point 2, both photons travel the same comoving path but with the reflected photon leading  the delayed gravitationally lensed photon in time by $T_p(t)=T_p(t_0)R(t)/R(t_0)$.\footnote{This is an approximation that requires $H_dT_p\ll 1$.} Because both photons had the same frequency $\nu_1$ and period $\delta t_1$ at $t_1$ the ratio of the lensed  photon's frequency at $t_2$ to the reflected photon's frequency at $t_2-T_p(t_2)$ is given by the inverse ratio of their periods, $\delta(t_2-T_p(t_2))/\delta t_2$.
Because all photons redden equally after  $t_2$  the ratio of the frequencies of lensed  to unlensed CMB photons at equal $t$ is \underbar{constant}  and is given by
\bea
\frac{\nu_{\rm sc}(t)}{\nu_{\td}(t)}
&=& \frac{R(t)}{R(t-T_p(t))}\times\frac{\nu_{\rm sc}(t)}{\nu_{\td}(t-T_p(t))} \cr
&=& \frac{R(t)}{R(t-T_p(t))}\times\frac{\delta(t-T_p(t))}{\delta t}.
\label{nu1}
\eea
This analytic expression can be evaluated at the cosmic time $t_d$ equivalent to the redshift $z_d$ of the deflector, giving
\be
\frac{\nu_{\rm sc}}{\nu_{\td}}=1+H_d\frac{\partial\,[(1+z_d) T_p(z_d)]}{\partial\, z_d}.
\label{nu2}
\ee
The quantity $T_p\equiv(1+z_d) T_p(z_d)$ is the potential part of the time delay as seen at the observer given in Eq.\,(\ref{Tp}), written as a function of the deflector's redshift.
When Eq.\,(\ref{nu2}) is used in Eq.\,(\ref{calT1}), Eq.\,(\ref{calT2}) results.
Equation (\ref{nu1}) can be interpreted as a product of a time-delay effect and a redshift effect  associated with the density evolution.
The evolutionary effect is produced by a time dependence in the projected mass fraction, see Eq.\,(\ref{calTE}), or equivalently in the relative mass density, see Eq.\,(\ref{Total}), and is hence associated with a time dependence of the Newtonian potential of the relative perturbation.

\section{The ISW/RS Effect for an Embedded Point Mass Lens}\label{append:static}

Because Eq.~(\ref{calT2}) is new and its derivation in Appendix A is subtle we wish to confirm its validity by comparing its prediction for the late-time ISW/RS effect caused by an embedded point mass with the prediction made by the conventional approach.
The conventional approach dates back to \cite{Rees68} and requires a detailed calculation of the photon's deflected orbit. Equation (2) of \cite{Dyer76} gives the  needed ratio of  observed frequencies of a lensed CMB photon $\nu_{\rm sc}$ to an identical unlensed CMB photon $\nu_{\td}$
\be
\frac{\nu_{\rm sc}}{\nu_{\td}}
=\frac{R(t_2)}{R(t_1)}\frac{\nu_{\rm sc}(t_2)}{\nu_{\rm sc}(t_1)},
\label{nu1B}
\ee
where all frequencies are measured by observers comoving with the background FLRW cosmology.
This equality is a result of both lensed and unlensed photons being identically redshifted after lensing (after the lensed photon exits at cosmic time $t_2$ with frequency $\nu_{\rm sc}(t_2)$, see point 2 of Fig.~\ref{fig:cheese}) and both photons having the same frequencies  prior to lensing which begins at $t_1$ with frequency $\nu_{\td}(t_1)=\nu_{\rm sc}(t_1).$

Using Eqs.~(9) and (10) of \cite{Kantowski10} and details of the photon's orbit that follow,  Eq.~(\ref{nu1B}) can be evaluated to give the ISW/RS effect for a flat background universe, but only after a significant amount of algebra.
The result is
\bea
\frac{{\cal T}+\Delta {\cal T}}{{\cal T}} &=&\frac{R(t_2)}{R(t_1)}\frac{\nu_{\rm sc}(t_2)}{\nu_{\rm sc}(t_1)}=\cr
&&1-\Bigg[\frac{\beta_1}{6}\frac{r_s}{r_0}\left(15\cos\pht+\cos 3\pht +12\log\tan\frac{\pht}{2}\right)\sin\pht  -\frac{2}{3}\frac{r_s}{r_0}\Lambda r_0^2\cos^3\pht\cot\pht\cr
&& -\frac{3}{4}\frac{r_s^2}{r^2_0}\sin^2\pht\cos\pht\left(7\cos\pht+\cos 3\pht+4\log\tan\frac{\pht}{2}\right)
\Bigg],
\label{nu2B}
\eea
where  the photon's orbit has been used to give its exit time $t_2$ and exit frequency $\nu_{\rm sc}(t_2)$ as a function of its entry angle $\pht$ and its minimum radial Kottler coordinate $r_0$.
The reader is directed to Fig.~1 of \cite{Kantowski10} for the definition of the entrance angle $\pht$ and minimal radial coordinate $r_0$, and to Eq.~(8) for a definition of the expansion rate $\beta_1$ of the lens void's boundary at entry point 1.
The reader is now in a position to appreciate the simplicity of using Eq.~(\ref{calT2}) rather than Eq.~(\ref{nu2B}) to evaluate the ISW effect.
The potential part of the time delay for the embedded point mass  is given in Eq.~(\ref{TPM})
from which a simple derivative with respect to $z_d$ in Eq.~(\ref{calT2}) gives Eq.~(\ref{calTT}) which is proportional to the time delay ${ T}_p^{{\rm PM}}(\theta_I/\theta_M)$ itself, i.e., for this lens
\be
\frac{\Delta{\cal T}}{{\cal T}}= \frac{\Delta{\cal T}_{\rm T}}{{\cal T}}=\frac{H_d\ T_p^{\rm PM}(\theta_I/\theta_M)}{1+z_d\ }.
\label{calTPM}
\ee
To compare Eq.~(\ref{nu2B}) with Eq.~(\ref{calTPM}) the Kottler impact variables $\beta_1$, $\pht,$ and $ r_0$ in Eq.~(\ref{nu2B}) must be converted to image positions $\theta_I$, expansion velocities $\beta_d$, lens sizes $r_d$, and maximum image positions $\theta_M$ all evaluated at fixed cosmic time $t_d$ defined by the deflector's redshift $1+z_d=R(t_0)/R(t_d)$.
The needed change of variables given in the next three equations can be found in Eq.~(3) of \cite{Kantowski13}, Eq.~(A1) of \cite{Kantowski13}, and Eqs.~(2) and (15) of \cite{Chen11}
\bea
\pht &=& \sin^{-1}\frac{\theta_I}{\theta_M}+\bdelta\beta_d\frac{\theta_I}{\theta_M} +{\cal O}(\bdelta^2),\cr
r_0&=&r_d\sin\pht(1-\bdelta\beta_d\cos\pht) +{\cal O}(\bdelta^2),\cr
\beta_1&=&\beta_d +\bdelta\left(\frac{r_s}{2r_d}-\frac{\Lambda r_d^2}{3}\right)\cos\pht +{\cal O}(\bdelta^3),
\eea
and results in Eqs.~(\ref{nu2B}) and (\ref{calTPM}) giving identical results for $\Delta{\cal T}/{\cal T}$ to order $\bdelta^1$.

The reader can easily see that because there is no time evolution of the projected mass fraction ($f(\theta_I/\theta_M)=1$) or equivalently in the density contrast ($\delta = -1$) for an embedded point mass, only the time delay part, Eq.~(\ref{calTT}), contributes to the ISW effect.


\begin{thebibliography}{breitestes Label}

\bibitem[Abazajian \etal(2009)]{Abazajian09}  Abazajian, K. N., Adelman-McCarthy, J. K., Ag$\rm\ddot{u}$eros, M. A, et al.\ 2009, \apjs, 182, 543

\bibitem[Bennett et al.(2003)]{Bennett03}  Bennett, C. L.,   Halpern, M.,  Hinshaw, G., et al. 2003, ApJS, 148, 1

\bibitem[Bertschinger(1985a)]{Bertschinger85a} Bertschinger, E. 1985a, \apjs, 58, 1

\bibitem[Bertschinger(1985b)]{Bertschinger85b} Bertschinger, E. 1985b, \apjs, 58, 39

\bibitem[Birkinshaw(1999)]{Birkinshaw99} Birkinshaw, M. 1999, Phys. Rep.,  310, 97

\bibitem[Blandford \& Narayan(1986)]{Blandford86} Blandford, R.,  \& Narayan, R. 1986,  \apj,  310, 568

\bibitem[Boughn \& Crittenden(2004)]{Boughn04} Boughn, S., \& Crittenden, R. 2004, \nat,  427, 45

\bibitem[Cai \etal(2010)]{Cai10}  Cai, Y-C., Cole, S., Jenkins, A., \& Frenk, C. S. 2010, \mnras,  407, 201

\bibitem[Cai \etal (2014)]{Cai14} Cai, Y-C., Neyrinck, M. C., Szapudi, I., et al.\  2014, \apj, 786, 110

\bibitem[Ceccarelli \etal (2013)]{Ceccarelli13} Ceccarelli, L., Paz, D., Lares, M.,  et al.\ 2013, \mnras, 434, 1435

\bibitem[Chen \etal (2010)]{Chen10} Chen, B., Kantowski, R., \&  Dai, X. 2010,  \prd,  82, 043005

\bibitem[Chen \etal (2011)]{Chen11}  Chen, B., Kantowski, R., \&  Dai, X. 2011,  \prd,  84, 083004

\bibitem[Chen \etal (2015)]{Chen13}  Chen, B., Kantowski, R., \&  Dai, X. 2015,  \apj, in press (arXiv1310.7574)

\bibitem[Cole \& Efstathiou(1989)]{Cole89} Cole, S., \& Efstathiou, G. 1989, \mnras, 239, 195

\bibitem[Cooke \& Kantowski(1975)]{Cooke75}  Cooke, J. H., \&  Kantowski, R. 1975, \apjl, 195, L11

\bibitem[Cooray (2002a)]{Cooray02a} Cooray, A. 2002a, \prd, 65, 083518

\bibitem[Cooray (2002b)]{Cooray02b} Cooray, A. 2002b, \prd, 65, 103510

\bibitem[Cooray \& Seto(2005)]{Cooray05} Cooray, A., \& Seto, N. 2005, \jcap, 0512, 004

\bibitem[de Lapparent \etal(1986)]{deLapparent86}  de Lapparent, V.,  Geller, M. J.,  Huchra, J. P. 1986, \apjl, 302, 1

\bibitem[Dyer(1976)]{Dyer76} Dyer, C. C. 1976, \mnras, 175, 429

\bibitem[Dup\'e et al.(2011)]{Dupe11} Dup\'e, F.-X., Rassat, A., Starck, J.-L., Fadili, M. J.\ 2011, A\&A, 534, 51

\bibitem[Einstein \& Straus(1945)]{Einstein45} Einstein, A., \&  Straus, E. G. 1945, Rev. Mod. Phys., 17, 120

\bibitem[Eisenstein et al.(2005)]{Eisenstein05} Eisenstein, D. J., Zehavi, I., Hogg, D. W., et al.\ 2005, \apj, 633, 560

\bibitem[Finelli \etal (2014)]{Finelli14} Finelli, F., Garc$\rm\acute{\i}$a-Bellido, J., Kov$\rm\acute{a}$cs, A., Paci, F., Szapudi, I. 2014, \mnras, submitted, arXiv1405.1555

\bibitem[Flender et al.(2013)]{Flender13} Flender, S., Hotchkiss, S., \& Nadathur, S. 2013, \jcap, 02, 013

\bibitem[Granett \etal(2008a)]{Granett08a}  Granett, B. R.,  Neyrinck, M. C., \&  Szapudi, I. 2008a, \apjl,  683, 99

\bibitem[Granett \etal(2008b)]{Granett08b}  Granett, B. R.,  Neyrinck, M. C., \&  Szapudi, I. 2008b, arXiv.0805.2974

\bibitem[Hamaus \etal (2014)]{Hamaus14} Hamaus, N., Sutter, P. M., Wandelt, B. D. 2014, \prl, 112, 251302

\bibitem[Hern\'andez-Monteagudo (2010)]{Hernandez10}  Hern\'andez-Monteagudo, C. 2010, A\&A, 520, 101

\bibitem[Hern\'andez-Monteagudo \& Smith (2013)]{Hernandez13}  Hern\'andez-Monteagudo, C., \& Smith R. E.\  2013, \mnras, 435, 1094

\bibitem[Ho \etal (2008)]{Ho08} Ho, S., Hirata, C.,  Padmanabhan, N., et al.\  2008, \prd,  78, 043519

\bibitem[Hotchkiss \etal (2015)]{Hotchkiss15} Hotchkiss, S., Nadathur, S., Gottl$\rm\ddot{o}$ber, S., et al.\ 2015, \mnras, 446, 1321

\bibitem[Hu(2000)]{Hu00} Hu, W.\ 2000, \apj, 529, 12

\bibitem[Ili\'c \etal (2013)]{Ilic13}  Ili\'c, S., Langer, M., \&  Douspis, M. 2013, A\&A, 556, 51

\bibitem[Inoue \& Silk(2006)]{Inoue06} Inoue, K. T., \&  Silk, J. 2006, \apj, 648, 23

\bibitem[Jarrett \etal (2000)]{Jarrett2000} Jarrett T. H., Chester T., Cutri R., et al.\ 2000, \aj, 119, 2498

\bibitem[Kantowski \etal (2010)]{Kantowski10} Kantowski, R., Chen, B., \&  Dai, X. 2010,  \apj, 718, 913

\bibitem[Kantowski \etal (2012)]{Kantowski12} Kantowski, R., Chen, B., \&  Dai, X. 2012, \prd,  86, 043009

\bibitem[Kantowski \etal (2013)]{Kantowski13}  Kantowski, R., Chen, B., \& Dai, X. 2013,  \prd, 88, 083001

\bibitem[Kantowski \etal (2014)]{Kantowski14}  Kantowski, R., Chen, B., \& Dai, X. 2014,  \prd, in press, arXiv1410.4608

\bibitem[Kim \etal (2013)]{Kim13}  Kim, J., Rotti, A., \& Komatsu, E. 2013, \jcap, 04, 021

\bibitem[Kottler (1918)]{Kottler18} Kottler, F.\ 1918, Ann. Phys., Lpz., 361, 401

\bibitem[Leclercq \etal (2014)]{Leclercq14} Leclercq, F., Jasche, J., Sutter, P. M., et al.\ 2014, arXiv1410.0355

\bibitem[Mart\'inez-Gonz\'alez \etal(1990)]{Martinez90} Mart\'inez-Gonz\'alez, E., Sanz, J. L., \&  Silk, J. 1990, \apjl,  355, L5

\bibitem[Maturi \etal (2007)]{Maturi07} Maturi, M., Dolag, K., Waelkens, A., et al.\ 2007, A\&A,  476, 83

\bibitem[Miralda-Escud\'e et al. (2000)]{Miralda00} Miralda-Escud\'e, J., Haehnelt, M., \& Rees, M. J. 2000, \apj, 530, 1

\bibitem[Nadathur \etal (2012)]{Nadathur12} Nadathur, S., Hotchkiss, S., \&  Sakar, S. 2012, \jcap, 06, 042

\bibitem[Nadathur \etal (2014)]{Nadathur14a} Nadathur, S., Lavinto, M., Hotchkiss, S., \& R$\rm\ddot{a}$s$\rm\ddot{a}$nen, S. 2014, \prd, 90, 103510

\bibitem[Nadathur \& Hotchkiss (2014)]{Nadathur14} Nadathur, S., Hotchkiss, S. 2014, \mnras, 440, 1248

\bibitem[Nottale(1984)]{Nottale84} Nottale, L. 1984, \mnras,  206, 713

\bibitem[Pan \etal (2012)]{Pan12} Pan, D. C.,  Vogeley, M. S.,  Hoyle, F., et al.\ 2012, MNRAS,  421, 926

\bibitem[Panek(1992)]{Panek92} Panek, M. 1992, \apj,  388, 225

\bibitem[Perlmutter \etal(1999)]{Perlmutter99} Perlmutter, S., Aldering, G., Goldhaber, G., et al.\ 1999, \apj, 517, 565


\bibitem[Planck Collaboration \etal (2014a)]{Planck14a} Planck Collaboration,  Ade, P. A. R.,  et al.\ 2014a, A\&A, 571, 24 

\bibitem[Planck Collaboration \etal (2014b)]{Planck14b}  Planck Collaboration,  Ade, P. A. R., et al.\ 2014b, A\&A, 571, 19 







\bibitem[Press \& Schechter(1974)]{Press74} Press, W. H.,  \& Schechter, P. 1974,  \apj,  187, 425

\bibitem[Rassat \etal (2007)]{Rassat07} Rassat, A., Land, K., Lahav, O., Abdalla, F. B. 2007, \mnras, 377, 1085

\bibitem[Rees \& Sciama (1968)]{Rees68}  Rees, M. J., \& Sciama, D. W. 1968, Nature,  217, 511

\bibitem[Riess \etal(1998)]{Riess98} Riess, A. G., Filippenko, A. V., Challis, P., et al.\ 1998, \aj, 116, 1009

\bibitem[Rudnick \etal(2007)]{Rudnick07} Rudnick, L., Brown, S. \&  Williams, L. R. 2007, \apj, 671, 40

\bibitem[Sachs \& Wolfe(1967)]{Sachs67}  Sachs, R. K., \& Wolfe, A. M. 1967, \apj, 147, 73

\bibitem[Sakai \& Inoue(2008)]{Sakai08} Sakai, N., \&  Inoue, K. T. 2008, \prd, 78, 063510

\bibitem[Schneider(1985)]{Schneider85} Schneider, P. 1985, A\&A, 143, 413

\bibitem[Schneider \etal(1992)]{Schneider92} Schneider, P.,  Ehlers, J., \& Falco, E. E. 1992, Gravitational Lenses (1st ed.; Berlin: Springer-Verlag).

\bibitem[Sch\"ucking(1954)]{Schucking54} Sch\"ucking, E. 1954,  Z. Phys., 137, 595

\bibitem[Seljak(1996a)]{Seljak96a} Seljak, U. 1996a, \apj,  463, 1

\bibitem[Seljak(1996b)]{Seljak96b} Seljak, U. 1996b, \apj,  460, 549

\bibitem[Sheth \& van de Weygaert(2004)]{Sheth04}  Sheth, R. K., \&  van de Weygaert, R. 2004, \mnras,   350, 517

\bibitem[Smith \etal (2009)]{Smith09} Smith, R. E., Hern$\rm\acute{a}$ndez-Monteagudo, C., Seljak, U. 2009, \prd, 80, 063528

\bibitem[Sunyaev \& Zeldovich(1980)]{Sunyaev80}  Sunyaev, R. A., \& Zeldovich,Y. B. 1980, \mnras, 190, 413

\bibitem[Sutter \etal (2012)]{Sutter12}  Sutter, P. M., Lavaux, G.,  Wandelt, B. D., \& Weinberg, D. H. 2012, \apj, 761, 44

\bibitem[Sutter \etal (2014)]{Sutter14}  Sutter, P. M., Lavaux, G.,  Wandelt, B. D., \& Weinberg, D. H., Warren, M. S. 2014, \mnras, 438, 3177

\bibitem[Szapudi \etal (2014a)]{Szapudi14a} Szapudi, I., Kov$\rm\acute{a}$cs, A., Cranett, B. R., et al.\ 2014, \mnras, submitted, arXiv1405.1566

\bibitem[Szapudi \etal (2014b)]{Szapudi14b} Szapudi, I., Kov$\rm\acute{a}$cs, A., Cranett, B. R., et al.\ 2014, arXiv1406.3622

\bibitem[Tegmark et al.\ (2003)]{Tegmark03} Tegmark, M., de Oliveira-Costa, A., \&  Hamilton, A. J. S. 2003, \prd, 68, 123523

\bibitem[Valkenburg(2009)]{Valkenburg09} Valkenburg, W. 2009, \jcap, 06, 010

\bibitem[Velva \etal (2004)]{Velva04}  Vielva, P.,  Mart\'inez-Gonz\'alez, E., Barreiro, R. B., et al.\ 2004, \apj, 609, 22

\bibitem[Vishniac(1987)]{Vishniac87} Vishniac, E. T. 1987, \apj, 322, 597

\bibitem[Watson \etal (2014)]{Watson14} Watson, W. A., Diego, J. M.,   Gottl$\rm\ddot{o}$ber, S., et al.\ 2014, \mnras, 438, 412

\bibitem[Zeldovich \& Sunyaev(1969)]{Zeldovich69}  Zeldovich, Y. B.,  \&  Sunyaev, R. A. 1969,  Ap\&SS, 4, 301

\end{thebibliography}
\end{document}